\documentclass[10pt,letterpaper,conference]{IEEEtran}

\usepackage{inconsolata}  
\usepackage{hyperref}     
\usepackage{breakurl}     
\usepackage{color}        

\hypersetup{
    pdfauthor={Andreas Tiemeyer, Tom Melham, Daniel Kroening, John O'Leary},   
    pdftitle={CREST: Hardware Formal Verification with \\ANSI-C Reference Specifications},   
    colorlinks=true,                          
    linkcolor=black,                          
    citecolor=blue,                           
    urlcolor=blue                             
}

\title{CREST: Hardware Formal Verification with \\ANSI-C Reference Specifications}
\date{\today}

\author{\IEEEauthorblockN{Andreas Tiemeyer, Tom Melham, Daniel Kroening}
        \IEEEauthorblockA{University of Oxford, Department of Computer Science\\
                          \{andyer, melham, kroening\}@cs.ox.ac.uk}
        \and
        \IEEEauthorblockN{John O'Leary}
        \IEEEauthorblockA{Intel Corporation\\ 
                          john.w.oleary@intel.com}}

\begin{document}

\maketitle
\thispagestyle{plain}
\pagestyle{plain}

\begin{abstract}
This paper presents CREST, a prototype front-end tool intended as an add-on to commercial EDA formal verifcation environments. CREST is an adaptation of the CBMC bounded model checker for C, an academic tool widely used in industry for software analysis and property verification. It leverages the capabilities of CBMC to process hardware datapath specifications written in arbitrary ANSI-C, without limiting restrictions to a synthesizable subset. We briefly sketch the architecture of our tool and show its use in a range of verification case studies.
\end{abstract}

\section{Introduction}

Schedule pressure demands continuous improvements to productivity in all semiconductor design phases.  Verification and test is often a critical bottleneck, and  a key factor is the choice of specification language.  For datapath designs, ANSI-C, C++ and SystemC are often the best languages to capture intent. Functional datapath specifications in these languages yield a higher level of abstraction, reduce code size, and speed up simulation and regression.

In this work, we focus on the scenario in which a designer maintains a high-level specification written in ANSI-C and the design under test is a low-level or synthesizable implementation in some standard HDL. In industrial practice, this scenario is often encountered with designs such as binary floating point units~\cite{Jones:2001:PFV,Mukherjee:2016:ECF} and GPUs.  Our work  is, however, also applicable to circuits with non-trivial control.

We have developed a tool for this setting called CREST, a prototype front-end intended as an add-on to commercial formal verification tools. It works by translating high-level ANSI-C specifications---potentially very large ones---into a low-level logical representation that is especially tuned to analysis by formal proof engines, such as SMT and SAT solvers. To be as general as possible, CREST uses very simple Verilog code as the intermediate language for communication of processed specifications to downstream EDA tools. The idea is to make high-level C specifications, expressed by almost arbitrary ANSI-C code, available for use as \textit{reference models} or \textit{specifications} in the formal verification of a RTL designs in an arbitrary commercial verification tool.

There are, of course, many existing compilers that map C to RTL, including several commercial C to RTL formal verification tools. There are also commercial and academic high-level synthesis tools that generate Verilog code---intended to represent a viable hardware circuit design---from higher-level C specifications. The novelty of our approach that we handle full ANSI-C, without artificial restriction to a synthesizable or other limited subset. This includes:

\begin{itemize}
\item arithmetic conversions, including conversions to and from floating-point types
\item typecasting, including typecasting to and from pointer types
\item deeply nested composite datatypes, such as structs of unions of structs
\item pointer dereferencing
\item passing of pointers as function arguments
\item dynamically-assigned function pointers
\end{itemize}

\noindent Our aim is to make it possible to integrate simulation models and other high-level models written in an unconstrained and natural style of ANSI-C into an RTL formal verification flow. 

\section{Architecture and Implementation}

As a software implementation, CREST is an adaptation of the CBMC bounded model checker for C~\cite{cbmc}, a substantial Oxford research tool developed by Professor Daniel Kroening that features a highly accurate, fully scalable bit-level logical semantics of full ANSI-C. CBMC is widely used by leading technology companies, including Bosch, General Electric, TATA and AWS~\cite{aws} and is also the core technology of Diffblue, an Oxford spinout company using AI to provide software synthesis and automated testing.

In CBMC, verification is done by unwinding program loops, symbolically executing the program, and passing the resulting equations to a decision procedure. The algorithms that do this have been highly engineered for efficiency, accuracy, and scalability through more than a decade of intensive use in industry and academic research.

Our solution exploits this capability by tapping into the CBMC internal data flow just after symbolic execution. From this representation, it then generates a low-level Verilog model that exactly captures the bit-level semantics of the effect of the symbolic program execution. This can then be read by a downstream RTL verification tool, including commercial tools from EDA vendors. It is important to note that the resulting Verilog is not intended to be a synthesised circuit design, but is instead a low-level representation that is tuned for analysis by SAT, SMT and other proof engines.

The result of CBMC's symbolic execution is a set of variable assignments that satisfy the single static assignment rule. All the high-level constructs of C will have been interpreted away, leaving a relatively simple representation of the program's semantics. Along the way, CBMC does a very significant amount of optimization and simplification---all aimed at optimising the results of symbolic simulation for processing by formal engines. CBMC also maintains full backannotation information, connecting each expression in the resulting equations to the file and line number of its C source. 

CREST then translates each variable in this representation to a Verilog bitvector and each C expression on the right-hand side of the assignments to a corresponding Verilog expression. Where possible, the
translation is done in the most straightforward way, mapping each C operator to an equivalent Verilog operator.  This is not, however, always possible. For example, Verilog does not allow the extraction of a bit range from an expression, but only from a variable. Our tool introduces auxiliary variables to solve this problem. CREST outputs the resulting Verilog code, and also makes available full backannotation information derived from CBMC.

The primary benefit of this approach is that it leverages CBMC's established and scalable capability to handle full ANSI-C, making a broad spectrum of high-level C models accessible as specifications for RTL verification.  

\section{Case Studies}

To give a flavour of the range of specifications that CREST can process, and the types of downstream verification enabled, we sketch five case studies using the tool.

\subsection{Softfloat vs VGM Floating Point Add}

SoftFloat~\cite{softfloat} is an architectural, IEEE-conformant software implementation in C of floating-point operations. It is owned and maintained by John Hauser (UC Berkeley) and widely considered a gold standard in high-performance software models of FP operations. CREST can take any SoftFloat interface function (e.g.~\texttt{f32{\_}add}, \texttt{f64{\_}mul}) as a starting point and generate Verilog that is equivalent to the function execution. This can then be used in a downstream EDA tool as a specification for equivalence verification against another reference model in RTL or, with appropriate datapath verification techniques deployed in the downstream tool~\cite{Seger:2005:IEE}, an RTL circuit implementation.

The SoftFloat function \texttt{f32{\_}add} implements 32 bit floating-point addition. To process this with CREST, we first create a C function that acts as a wrapper around the SoftFloat function and also defines the inputs and outputs of the resulting Verilog module. This function looks as follows:

\begin{verbatim}
  #include <SoftFloat.h>
  void f32_add_wrapper()
  {
    uint32_t x, y;
    C2V_SAMPLE_INPUT(uint32_t, x);
    C2V_SAMPLE_INPUT(uint32_t, y);
 
    float32_t xf, yf, rf;
    xf.v = x;
    yf.v = y;
    rf = f32_add(xf, yf);
    uint32_t res = rf.v;
 
    C2V_DRIVE_OUTPUT(uint32_t, res);
  }
\end{verbatim}

\noindent The macros \texttt{C2V{\_}SAMPLE{\_}INPUT} and \texttt{C2V{\_}DRIVE{\_}OUTPUT} define the interface of the Verilog module created by our tool. Its inputs are bitvectors of width 32. These are interpreted as floating point representations, not as integers, and passed to the SoftFloat function \texttt{f32{\_}add}. We then use the bit pattern of the result as the output of the Verilog module. 

From this C code, CREST generates a Verilog module with the following interface:

\begin{verbatim}
  module f32_add(
    input logic unsigned [31:0] x,
    input logic unsigned [31:0] y,
    output logic unsigned [31:0] res
  );
\end{verbatim}

\noindent The logic of the module is semantically equal to the C code of \texttt{f32{\_}add{\_}wrapper}, which calls SoftFloat's \texttt{f32{\_}add}  function.

This generated Verilog can now be compared by standard equivalence checking tools to an existing RTL reference model or  implementation of the same 32-bit floating point operation. In our work, we are benchmarking with the reference models provided by the Verilog Golden Model library (VGM). This is a high-quality reference library for floating point operations created by Warren Ferguson and Flemming Andersen and available under an open-source license. It implements standard arithmetic operations and fused multiply-add, and we have added a certain number of conversion operations. The VGM modules are parameterized in terms of the widths of the exponent and significand of floating-point values, so the library covers more than the standard IEEE precisions. It supports all conventional rounding modes and some of the recently-introduced non-IEEE rounding modes. Customization options allow for the definition of architecture-specific handling of unspecified behaviour involving NaNs.

Using any off-the-shelf sequential equivalence checking tool, it is straightforward to verify the equivalence of the floating point addition reference models of SoftFloat and VGM. The verifications go through fully automatically for half-, single- and double-precision. In academic experiments at Oxford with the SEC App in JasperGold version v2018.12, the runtimes are all very reasonable:

\medskip

\begin{tabular}{@{}r|r@{}} 
\multicolumn{1}{l|}{\textbf{Width}} & 
\multicolumn{1}{l}{\textbf{Runtime}} \\ 
\hline
16 bits & 6.11 sec. \\
32 bits & 59.68 sec. \\
64 bits & 2068.38 sec. \\
\end{tabular}

\medskip

\noindent These runtimes are for a 3.07\,Ghz 4-core Intel Xeon X5667 machine with 48\,GB memory running Linux kernel 4.13.

\subsection{Floating Point Multiplication}

Equivalence of the SoftFloat and VGM reference models for floating
point addition can be verified end-to-end by a state of the art
equivalence verification tool. But for floating point multiplication, 
a problem decomposition is needed. This is very typical of challenging
data path proofs~\cite{OLeary:2013:RST, Mukherjee:2016:ECF, Jones:2001:PFV}.

For multiplication, SoftFloat and VGM take different approaches to handling subnormal
operands. SoftFloat normalizes each operand prior to performing an integer multiplication of the significands.
VGM performs the multiplication directly on the (possibly subnormal) significands and applies a corrective
shift to the product afterwards. Furthermore, the widths of the integers holding the significands
differ between the two models: 64 bits in SoftFloat (modelling C's '{\tt *}' operator) versus 48 bits in VGM. 
These differences make automatic equivalence verification difficult, but they can easily be tackled
with an appropriate proof decomposition.  

The decomposition is done in the equivalence checking step between VGM and CREST-generated
Verilog from the SoftFloat model. The following cases must be considered:

\begin{enumerate}
\item Special operands: infinities, NaNs, and the like.
\item Neither operand is special and at least one operand is zero.
\item Both operands are non-zero normal or subnormal floats.
\end{enumerate}

\noindent Equivalence of the multipliers in cases 1 and 2 is easily proved automatically in an 
RTL formal verification tool, as is the exhaustiveness of this case analysis.

Equivalence of the models in case 3 requires establishing some
invariants linking the models. In the subcase when both operands are
normal floats, we first prove correspondences between the fraction
fields of the VGM operands and inputs to SoftFloat's significand
multiplier:

\begin{verbatim}
  SoftFloat.mul_in_1[63:31] = 33'b0
  SoftFloat.mul_in_1[30:7]  = {1'b1,Vgm.fp1[22:0]} 
  SoftFloat.mul_in_1[6:0]   = 7'b0
  SoftFloat.mul_in_2[63:32] = 32'b0
  SoftFloat.mul_in_2[31:8]  = {1'b1,Vgm.fp2[22:0]}
  SoftFloat.mul_in_2[7:0]   = 8'b0
\end{verbatim}

\noindent Assuming these correspondences, we then establish the correspondence between the significands resulting from
integer multiplications in the two models:

\begin{verbatim}
  SoftFloat.mul_out[63]}   = 1'b0
  SoftFloat.mul_out[62:15] = vgm.mul_out[47:0]
  SoftFloat.mul_out[14:0]  = 15'b0
\end{verbatim}

\noindent Finally, we complete the proof of the subcase by proving that
SoftFloat and VGM compute identical products, assuming correspondence
of their multiplications. The remaining subcases, when one or
both operands are subnormal, require additional case splitting based
on the magnitude of the corrective shift applied in the VGM.

The case analyses, including proofs of each property and
exhaustiveness of each case split, were scripted in Tcl and executed
in a standard EDA vendor formal verification tool.

\subsection{Approximate Reciprocal}

The C model in this case study is publicly distributed by
Intel~\cite{rcp14}. It is a software model of the behaviour of the VRCP14\{P,S\}S
machine instructions, members of the Intel$^\textrm{\small\textregistered}$ AVX-512
instruction family that compute approximate reciprocals of single
precision floating point values, with relative error of less than
\(2^{-14}\). The code was written by numerics architects without
consideration of the needs of high level synthesis or formal
verification. This code is known to be problematic for existing C to
Verilog solutions, because it passes pointers as function arguments
and employs various `type punning' techniques such as the use of
typecasting to access bit patterns in a float variable:
\begin{verbatim}
  float x; 
  int *xp = (int*)&x;
  *xp = *xp & 0x003fffff
\end{verbatim}
CBMC handles such tricks with aplomb, though they are often not
accepted as within the synthesizable subset by commercial C to
Verilog solutions.

Another challenge arose because the authors of the C model wrote the
code recursively. The reciprocal approximation is computed by table
lookup and interpolation, and the approximation requires that the
approximand lies in the interval
\([1.0,2.0)\).
Other floating point inputs are first scaled to lie within the
interval, the approximation function is called recursively on the
scaled float, and the result of the recursive call is `de-scaled'
accordingly. We chose to unroll the recursion by hand, and used CBMC
as a model checker to prove the assertions that justify the
correctness of the unrolling transformation.

The Verilog code specification generated by CREST has been
successfully checked against a proprietary, hand-coded reference and
against an optimised RTL design from a shipping microprocessor, using
both EDA vendor tooling and proprietary symbolic simulation engines.

\subsection{Google's WebM VP9 Codec} 

We translate an implementation of matrix transformations used in the WebM VP9 encoder implementation by Google~\cite{webm:libvpx}. The transformations act on a 16 $\times$ 16 matrix of bits represented as an int16{\_}t array of size 16. Each transformation is represented by a tuple of function pointers, which implement the one-dimensional transforms acting on matrix rows and columns. A primary input to the top level transformation function is an indicator variable that selects the specific transformation to be used. In order to translate this code to Verilog, a tool must be able to handle function pointers that cannot be resolved statically.

The code also contains several user-written assertions, which are are translated into SVA assertions by our tool. These assertions are located in a low-level function that implements a transformation done in two passes, and state that after the first pass an intermediate result has been calculated. We formally verify these assertions in the original C code using the native CBMC prover. We also verify the translation of the assertions in the generated Verilog using a leading commercial RTL verification tool.

\subsection{Sequential Floating-Point Adder} 

To demonstrate the ability to use specifications generated by our tool for verification of sequential designs, we take SoftFloat's \texttt{f32{\_}add} function as the specification for a synthesizable implementation of a 32-bit floating point adder that is optimized for area~\cite{dawson}.
The circuit implementation executes add operations one after the other, not in parallel. The next execution starts only after the previous execution has finished. Internally, it uses a finite state machine to define the current execution stage of the add operation. The design is clocked and changes of the execution stage happen on the rising edge.

\smallskip
The execution stages are: 
\begin{itemize}
\item \texttt{get}: read and store inputs
\item \texttt{unpack}: unpack the inputs into sign, exponent and significand
\item \texttt{special{\_}cases}: handle the special cases of inputs being not a number (NaN), infinity or zero
\item \texttt{align}: align significands to the larger exponent
\item \texttt{add}: add significands and set rounding bits
\item \texttt{normalize}: normalize the result
\item \texttt{round}: round the result to nearest even
\item \texttt{pack}: pack sign, exponent and significand of the result into a bitvectors
\item \texttt{put}: write the result to output
\end{itemize}

The implementation executes one operation per clock cycle. This includes the shifts during normalization and alignment, and as a result a single add operation can take up to 108 clock cycles. The implementation is therefore very different in structure to the Verilog generated by our tool from the SoftFloat C specification, which is combinatorial. Nevertheless we are able to show automatically, using a leading formal verification tool, that the implementation is standard compliant if and only if the specification is standard compliant.

\section{Conclusion: Benefits and Prospects}

Building the CREST front-end on CBMC brings two substantial benefits: a precise, bit-level interpretation of the semantics of full ANSI-C, and a built-in best-in-class model checker for C assertions. This enables us to support several usage scenarios aimed at RTL design verification against high-level C reference models.

\medskip

\noindent \textbf{RTL verification against ANSI-C specifications.}
This is the primary usage scenario: a C reference specification is translated into its low-level semantics, in a form tuned to processing by SAT and SMT engines. This is made available as very simple Verilog code, so that it can be used as a specification of intended behaviour for verification of a circuit using any downstream EDA verification tool.

\medskip

\noindent \textbf{Independent verification of HLS.} CREST provides a C to RTL path that could be used as an independent check on the results of high level synthesis (HLS). Starting from the same C source, an RTL circuit can be generated by HLS and a reference specification in Verilog generated by CREST. Comparing the two by equivalence checking would provide additional confidence in the correctness of circuit synthesis.

\medskip

\noindent \textbf{Proving properties about C specifications.}  Since
CBMC is part of our solution, its model checking capability can be
used to verify properties of the reference specification. This
includes user written assertions as well as pointer safety and buffer
overflow checks.  We exploited this capability to justify source
transformations in the approximate reciprocal example. Our tool
translates C assertions into SVA, so it is also possible to verify them
with the verification engines of a downstream EDA tool.

\medskip

\noindent \textbf{Using established C properties as helper assertions.}
Once properties of a C specification have been established by CBMC, they can in principle be used as helper assertions in downstream formal verification of a circuit against the specification. The hope is that the relatively high level of the properties that can be proved in C would, when translated by CREST, give the downstream verification engines helpful information. Experiments are needed to determine how useful this can be in practice.

\medskip

Current work on CREST is focussed on further case studies, including industrial C specifications, as well as exploring ways to extending support for C++ beyond the current capabilities of CBMC for specifications in that language.

\bibliographystyle{IEEEtran}
\bibliography{Tiemeyer-2019-CHF}

\end{document}